\documentstyle[12pt,psfig]{article}
\begin{document} 

\title{Pattern Formation in a 2D Elastic Solid} 

\author{A. C. E. Reid\\
Dept. of Materials Science and Engineering, \\ 
Northwestern University,\\ Evanston, Illinois USA 60208 
\and
R. J. Gooding\\
Dept. of Physics, Queen's University, \\ 
Kingston, Ontario Canada K7L 3N6}

\maketitle

\begin{abstract}
We present a dynamical theory of a two--dimensional martensitic
transition in an elastic solid, connecting a high--temperature
phase which is nondegenerate and has triangular symmetry, and a
low--temperature phase which is triply degenerate and has oblique symmetry.
A global mode--based Galerkin method is employed to integrate
the deterministic equation
of motion, the latter of which is derived by the variational principle
from a nonlinear, nonlocal Ginzburg--Landau theory which includes the
sound--wave viscosity. Our results display (i) the phenomenon of surface 
nucleation, and (ii) the dynamical selection of a length scale of the resultant
patterns.
\end{abstract}

Dynamical pattern formation, while canonically associated with fluid systems 
such as convection cells, or initially fluid systems, such as is the 
case in eutectic solidification, also occurs in fully coherent 
solid--to--solid structural phase transformations which are purely
elastic in character.  Such phase transformations, generally
known as ``Martensites'', involve the formation of regions of the
solid which become permanently strained relative to their undistorted
parent phase, without the diffusive transport of matter, and without
the introduction of crystallographic dislocations into the system.

A two--dimensional model of such a system is the principal topic of
this paper and is described in more detail below, but 
it is worth nothing that considerable insight was gained into 
this problem by the study of conceptually straightforward
one--dimensional models.  The work of Bales and Gooding [1]
dramatically illustrates the role of inertia in elastic
systems with degenerate product states.  This model included all
of the important features of the potential energy which are present
in the current study, and gave 
rise to a nontrivial, long--lived, patterned state.  This early 
work did not examine the role of the boundaries of
the system, but related work [2] using a single--product--state
potential and again using inertial Hamiltonian dynamics demonstrated
the role of the system boundary as a potential nucleating defect ---
in this system, a decaying subcritical excitation interacted with the
free boundary to drive the system into the product state well.  Because
of the nature of the potential energy of this system (a nondegenerate
product state), pattern formation was not possible. An extensive summary
of the 1D work carried out by the Queen's group can be found in Ref. [3].

In this paper we present a dynamical model of a $D > 1$ martensitic transformation. 
We study the particular example of the transition undergone
by a 2D elastic solid that is reminiscent of the 3D transition undergone
by a real material. In 2D we let the high--temperature phase be
nondegenerate and have triangular symmetry, and the low--temperature 
phase be triply degenerate and have oblique symmetry. This particular 
choice of parent and product phase symmetries leads to intriguing
self--similar patterns, some of which we hoped to be able to access dynamically. 
The equation of motion is formulated using a variational continuum mechanics 
approach, and the elastic displacement field evolves 
in time under deterministic, inertial Hamiltonian dynamics.
As the system evolves, it is forced to choose between the energetically 
equivalent minima, resulting in the formation of patterned states in which 
some of each possible product phase is present --- we enforce the absence of
dislocations by integrating the displacement field, not the components
of the strain tensor, thus ensuring that the elastic compatibility relations
are satisfied.  The system is observed to begin the displacement from the 
free boundaries, consistent with the critical nucleus of the system existing 
at the surface, and the hydrodynamic selection of a characteristic length-scale 
for the resulting patterns is demonstrated.

The real material [4] that inspired our two--dimensional model 
system is lead orthovanadate, ${\rm Pb_3(VO_4)_2}$, where 
a nested self--similar star--like patterned morphology occurs.
We take as the plane of the model system the plane normal to the long
axis of the trigonal unit cell in the physical system, and require
that the potential energy reproduce the corresponding three--fold
rotational symmetry.  In ${\rm Pb_3(VO_4)_2}$, the
structural transformation changes the trigonal parent--phase
unit cells to monoclinic product--phase cells, 
breaking the three--fold rotational symmetry along one of three equivalent 
directions.  We model this by including three degenerate strain minima 
in a Ginzburg--Landau potential, the minima corresponding to area--preserving
shear strains which locally stretch the system along one of three
equivalent lines spaced $120^\circ$ apart.  As in the one--dimensional
systems, we include dissipation, inertia, and strain--gradient
terms.

The dynamical variables of the model are the two components of the
continuum vector displacement field, $u_x(x,y)$ and $u_y(x,y)$.
Any value of the symmetric strain tensor for the system, derived
from these displacements, can be
expressed in terms of the three principal strains, which we
label $e_1$, $e_2$, and $e_3$, where the $e_1$ strain corresponds
to a uniform dilation of the system, and $e_2$ and $e_3$ correspond
to the two area--preserving shear strains which stretch the system 
along one axis and contract it along the normal to this axis.
The $e_2$ and $e_3$ strains are distinguished by having their 
stretch axes at $45^\circ$ to each other.

We implement a phenomenological free energy density which governs the 
dynamics as a temperature--dependent mechanical potential energy density.
This free energy has several components: The local portion
provides the essential nonlinearity of the system --- it is
this density, a function only of the principal strains, which
respects the three--fold symmetry of the system and which 
gives rise to the symmetry--breaking degenerate stable minima.
It is given by 
\begin{equation}
{\cal V}_{\rm local} = {1 \over 2}A{e_1}^2 + {1 \over 2}B({e_2}^2+{e_3}^2)
- {1 \over 3}C({e_2}^3-3e_2{e_3}^2)+{1 \over 4}D({e_2}^2+{e_3}^2)^2~.~
\label{eq:vlocal}
\end{equation}
The parameters $A$ and $B$ are the linear elastic constants of
the system, and vary with temperature.
To this linear order, the potential is elastically
isotropic.  The parameters $C$ and $D$ reflect the nonlinearity
of the system.  The parameter $D$ must be positive for the potential
to be globally stable, and the parameter $C$ governs the details
of the potential wells.  Changing the sign of $C$ changes the 
position of the global mimima in the $e_2$--$e_3$ plane, but not
their overall structure.

\begin{figure}
\hfil
\psfig{file=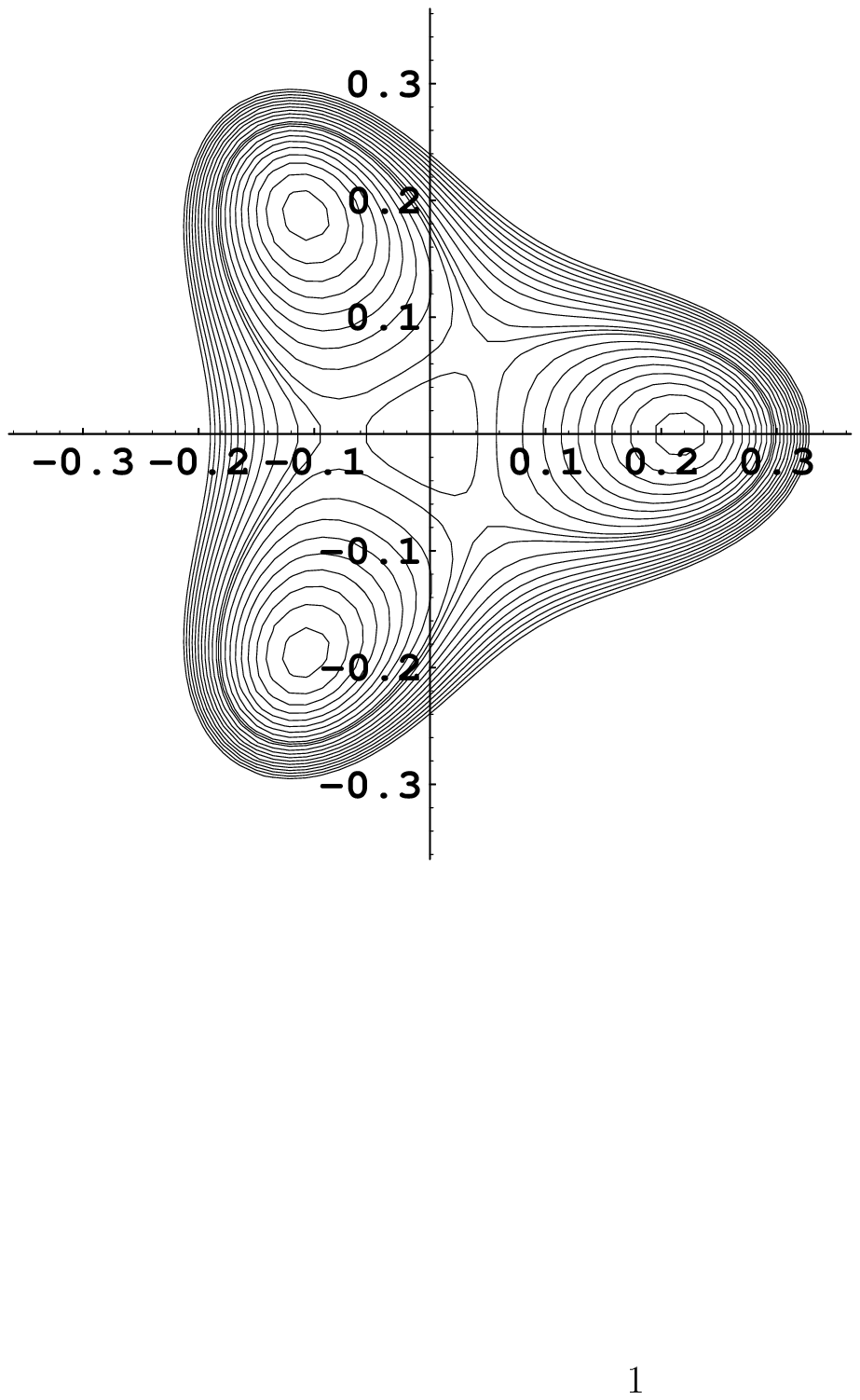,height=10.0cm}
\caption{Contour plot of the local potential energy, indicating
the central metastable well and the three stable wells corresponding
to strained states.}
\label{fig:cplot}
\end{figure}

This potential is illustrated in Figure 1, which 
is a contour--plot of the $e_2$--$e_3$ portion of Eq.~(\ref{eq:vlocal}).
The central metastable well is evident at the origin of the figure,
where $e_2=e_3=0$, and the three globally stable product wells
are evenly spaced at the same distance from the origin, at the
linear combinations of $e_2$ and $e_3$ corresponding to stretching
along the three $120^\circ$ lines.  It is worth noting at this point
that Fig. 1 is in the {\em strain} space of the system,
and that the positions of the product--state wells in this space
is not directly related to the orientation of the principal 
directions of expansion and compression for the corresponding strains.

In addition to the local terms, the correct description of the 
inertial dynamics requires that we also account for the kinetic
energy of the system, the nonlocal or Ginzburg terms of the 
potential, and the dissipation.  The Ginzburg terms prevent the 
system from forming unphysically small domains of the product state
by breaking the scale--invariance of the linear system, and the 
dissipative terms provide for the removal of kinetic energy from
the system as it goes from the initial state of high potential
energy to the final, patterned state of low potential energy.
The relevant kinetic, nonlocal, and dissipative terms are given by
\begin{eqnarray}
{\cal T}&=&{1\over 2}\rho ({\dot{u}_x}^2+{\dot{u}_y}^2)~~,\cr
{\cal V}_{\rm nonlocal}&=&{1 \over 2}g_1({e_{1,x}}^2+{e_{1,y}}^2)
+{1 \over 2}g_2({e_{2,x}}^2+{e_{2,y}}^2+{e_{3,x}}^2+{e_{3,y}}^2)~~,\cr
{\rm and}\quad{\cal R}&=&{1 \over 2}\gamma_{A}{\dot{e}_1}^2
   +{1 \over 2}\gamma_{B}({\dot{e}_2}^2+{\dot{e}_3}^2)~~.
\end{eqnarray}

All of the foregoing are intensive aspects of the system, independent
of the details of the system geometry.  The final ingredient in 
constructing the dynamics is to specify the system geometry.
We choose a diamond--shaped system with two $120^\circ$ corners and
two $60^\circ$ corners, as illustrated in the time--slices of
Figures 2 $\rightarrow$ 4.
The apparent stepped structure of the left--hand and right--hand
edges is due to the shading algorithm.  The model system 
has straight sides.
The system geometry reflects a pragmatic compromise
between the $120^\circ$ symmetry of the potential and the
computational tractability of a square system, to which this
geometry can be mapped for the purpose of performing the 
required numerical integrations.

The equation of motion for the dynamics of
this system is the usual inertial Hamiltonian dynamics for 
the displacement field.  The Lagrangian of the system is simply
the integral of the Lagrangian density obtained by subtracting
all of the potential--energy terms from the kinetic terms
in the usual way.  We have that 
\begin{equation}
L=\int\!\int ({\cal T}-{\cal V}_{\rm local}-{\cal V}_{\rm nonlocal})\cdot da~~,
\label{eq:lag}
\end{equation}
and the corresponding dissipation function is given by the 
area integral of the dissipation density,
\begin{equation}
R=\int\!\int {\cal R}\cdot da~~.
\label{eq:dis}
\end{equation}

The Lagrangian and dissipation function so obtained contain the
dynamical fields $u_x$ and $u_y$ through their contribution to the
strains which appear in the potentials, as well as explicitly in the 
kinetic energy density.  The equation of motion for the system
is obtained by functional differentiation of the Lagrangian and 
dissipation function with respect to the dynamical fields
and their time--derivatives, respectively, in the usual
way [5].  Abstractly, we simply have that
\begin{equation}
{\delta L \over \delta u}={\delta R \over \delta \dot{u}}~~.
\label{eq:eom}
\end{equation}

Numerically, there are two approaches which suggest themselves
at this point.  The first of these is to obtain an analytic 
equation of motion for the system from Eq.~(\ref{eq:eom}), and
then attempt to represent the dynamical fields and their
derivatives either as an expansion in some set of basis functions,
or on some kind of finite-element grid, and then deduce from the
analytic equation of motion a corresponding equation of 
motion for the model degrees of freedom.  This approach has
the considerable advantage of conceptual straightforwardness,
and was used with considerable success in the one--dimensional
models previously cited.
It is, however, complicated by the necessity of representing fourth
derivatives of the displacement field, which occur as a result
of the presence of the gradient terms, a complication which is 
rendered more difficult for the combinatorially large number
of mixed partials which occur in the two--dimensional system.

For this reason, we chose to pursue a slightly different 
approach, which involves the construction not of an approximated
equation of motion, but an approximated Lagrangian and dissipation
function.  The dynamical fields $u_x$ and $u_y$ are first 
expressed as expansions over a set of basis functions, in principle
complete but of necessity truncated, and the new Lagrangian and
dissipation function are constructed dependent on the amplitude
coefficients and their time--derivatives.  The solution to the
system is then the equation of motion for the coefficient amplitudes.
This is a globally--based modification of the familiar Galerkin method
for solving partial differential equations.

Formally, the two methods are equivalent, but practically speaking,
the latter method enjoys significantly better numerical stability,
and a small advantage in the speed with which solutions are obtained,
for a variety of reasons.  In the latter method, the spatial integrations
of the various densities need be computed only once, and lead to 
mode--coupling coefficients in the resulting mode--based Lagrangian.
The integrations involve products of lower--order derivatives of the
displacement fields, and in addition to the greater numerical stability
of this process, also lends itself to the exact computation of the
mode--coupling coefficients.  There need be no other approximations
in the system other than the necessary truncation of the set of 
modes used to represent the displacement fields. A complete discussion
of this technique will be given in a later publication.

The results of one particular dynamical run, illustrative
of all of the major effects observed in this study, are
presented in Figures 2 $\rightarrow$ 4.
Each panel of each figure is a snapshot of the $e_2$--$e_3$ strain
state of the system.  The panels of the first two figures,
Figs. 2 and 3, are 
evenly spaced in time and cover the early portion of the dynamics
through to approximately three times the speed--of--sound travel
time across the long axis of the fully--transformed system.
The final figure, Fig. 4, consists of snapshots
spaced at eight times the interval of the first two figures,
and shows the late stage of the dynamics.
In all of these figures, regions in which a given stable strain 
state is approximately
fully developed are indicated by the grey zones in these
figures.  The accompanying arrows indicate the direction in which
the region is stretched, although it should be kept in mind
that the minima correspond to area--preserving shear strains with
a corresponding contraction at $90^\circ$ to the expansion direction.
The black regions of these figures
correspond to portions of the system which are close
to the parent phase, with approximately zero strain.

\begin{figure}
\hfil
\psfig{file=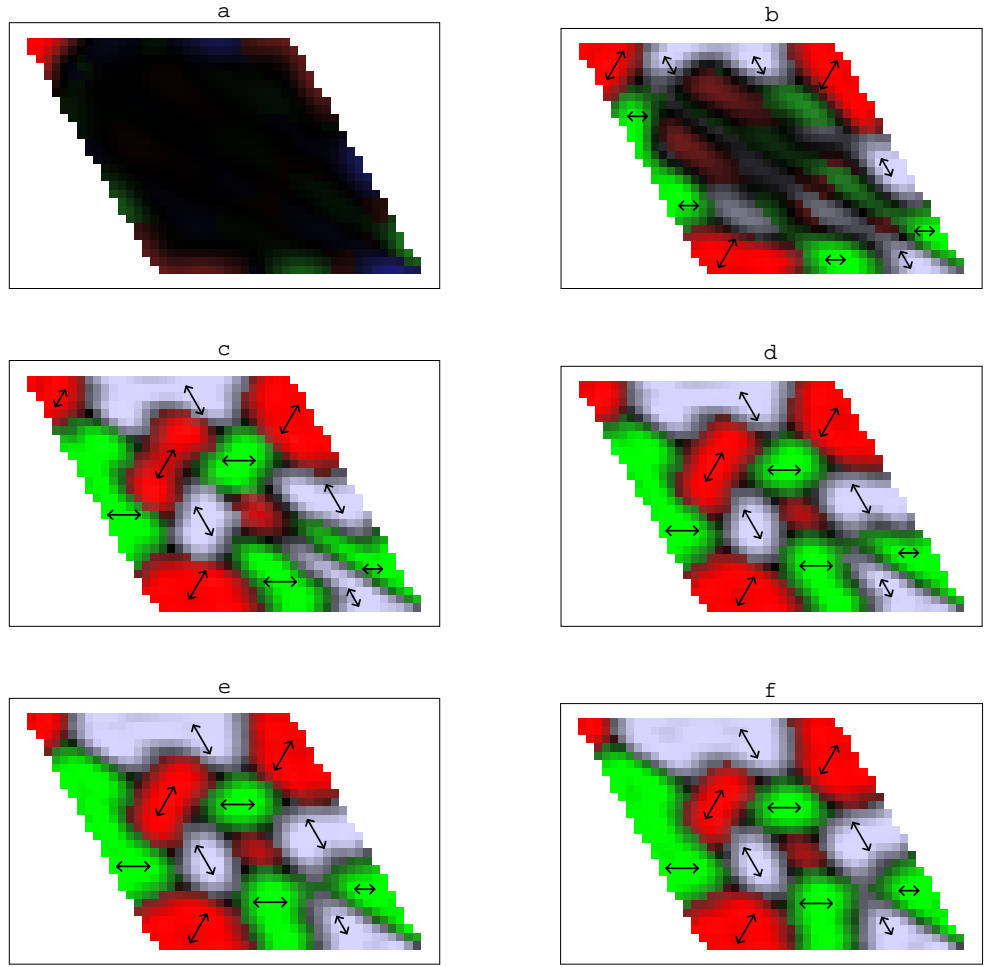,height=14cm}
\caption{Time--slices of the $e_2$--$e_3$ state of the dynamical
system.  The shaded grey regions in each panel are regions of the
system which have achieved different stable transformation strains,
the precise orientation being indicated by the accompanying arrows.
This figure shows a sequence of snapshots of the system during the
earliest stages of the transformation, during which transformed
domains initially form around the boundaries of the system, followed
by the formation of interior domains of approximately uniform 
size with complementary strain orientations.}
\label{fig:symd0}
\end{figure}

We present one particular study undertaken using
the mode--based dynamical technique described previously.
In order to expedite the dynamics, this study was
undertaken with the linear restoring force for shear,
coefficient $B$ of Eq.~(\ref{eq:vlocal}), strongly negative.
We began with low--amplitude random initial 
condition for the initial displacements, and with the initial
velocities set to zero, and proceeded under deterministic
inertial dynamics with free boundary conditions.  

The precise parameter values were chosen to be
$A=0.0$, $B=-2000$, $C=16547.7$, $D=118331.0$, $\gamma_A=0.1$,
$\gamma_B=6.0$, and $g_1=1.2$, and $g_2=4.0$.  This parameter
set gave product--state well depths of order one, in the arbitrary
but self--consistent energy units defined by the potentials, and
strain amplitudes for the well bottoms of $0.1$.

As indicated in Fig. 2, the system
begins the transformation from
the boundaries, first falling in to the stable minima at the
corners,
then subsequently along the edges, and then forming a patchwork
of domains throughout the 
system. Given the absence of linear restoring forces for the shear
strains, it is
noteworthy that this initial development stage for product
states does not take place immediately.  Local potential energy
considerations would suggest that the behaviour should be 
that of spinodal decomposition.  In this system, however, there are two
additional complementary effects, those of inertia and elastic
accommodation, which prevent the system from spontaneously 
transforming itself entirely into the product--phase, and both of these effects
account for why the transformation begins at the surface.

\begin{figure}
\hfil
\psfig{file=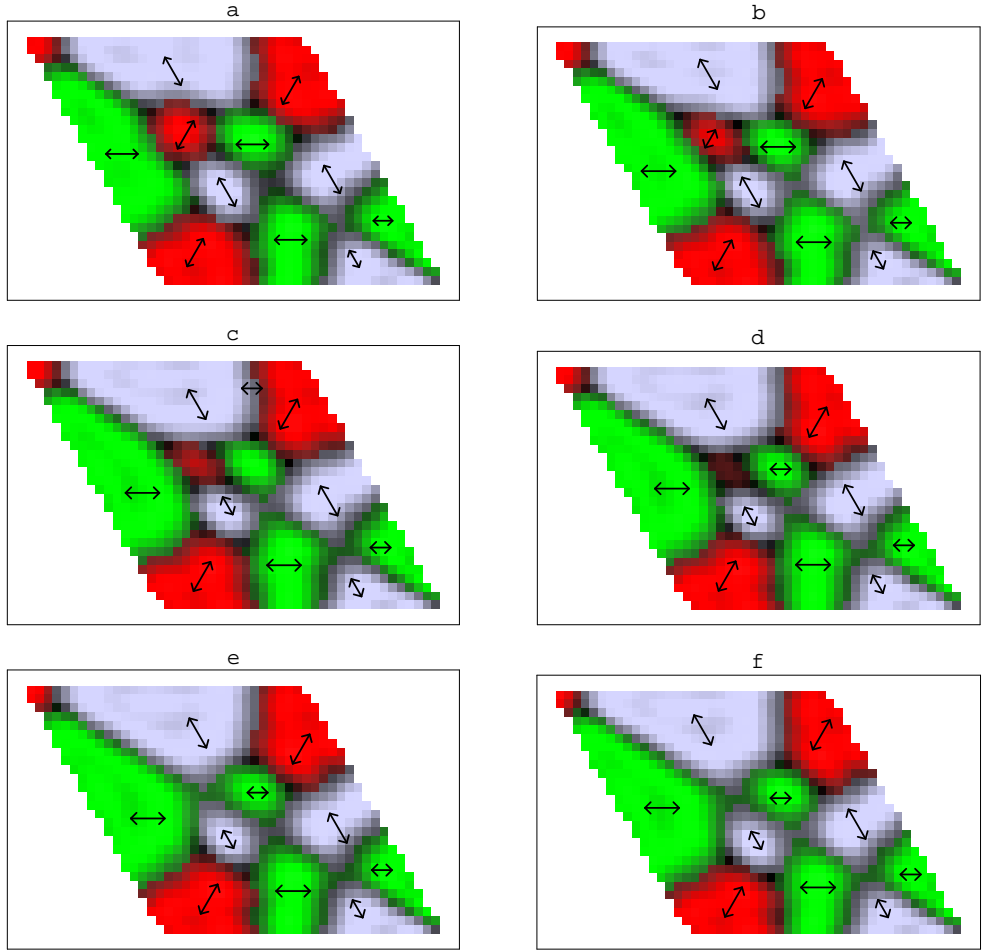,height=14cm}
\caption{Late--stage snapshots of the dynamical system.
This figure is a continuation of Fig. 3, but with the panels
separated by a time interval eight times larger than that of
Figs. 2 and 3.  The two complementarily--oriented domains in the
upper left portion of the figure expand and push the smaller domains
off the right--hand end, until finally the system consists of an 
arbitrarily long--lived twinned state, higher in energy than
the system's ground state.}
\label{fig:symd2}
\end{figure}


Elastic accommodation provides a penalty, in terms of elastic
energy, for forming an isolated region of product--phase in the
centre of the system.  In order to maintain the elastic 
coherency of the system, such a structure must be surrounded
by a displacement field which relaxes towards the unstrained
parent phase at distances large compared to the size of the
structure.  This accommodation effect is one of the principal
differences between elastic hydrodynamics and the more conventional
convective hydrodynamics, which, even in the inertial regime, 
is not subject to an analogous coherency constraint.

The accommodation constraint leaves open the possibility that
the system may suddenly form a system--wide network of 
self--accommodating domains.  Such a structure would certainly
rapidly minimize the potential energy of the system, but 
in this case it is the inertia of the system which prevents
it from following such a dynamical path.  In order to collectively
move all the parts of the system in the required co-ordinated 
manner would require a substantial amount of kinetic energy.
The energy required to support the required velocity field is
simply not available, and so we see instead that for both of
these reasons, the dynamics begin around the edges and the 
corners, where the accommodation constraint can be avoided
entirely and the inertia of the portion of the system near
the tip of one of the points is small.

Following the initial ``nucleation'' (though we use this word
with some caution) stage of the dynamics, the system 
forms an interior patchwork of transformed domains of 
approximately uniform size, about which we shall have
more to say in a moment.  The system then undergoes 
a coarsening process during which large domains tend to 
become larger, and smaller domains are absorbed into larger
domains or are expelled from 
the system, illustrated in Figs. 2 and 3.
The strain order parameter is not 
conserved, but large domains tend to grow at the expense of
smaller adjacent domains, and because of the inertial and accommodation
effects, these adjacent domains tend to consist of complementary product
phase.  Thus the dynamics give the appearance of the order
parameter being, in some sense, approximately 
conserved during this intermediate phase. 

\begin{figure}
\hfil
\psfig{file=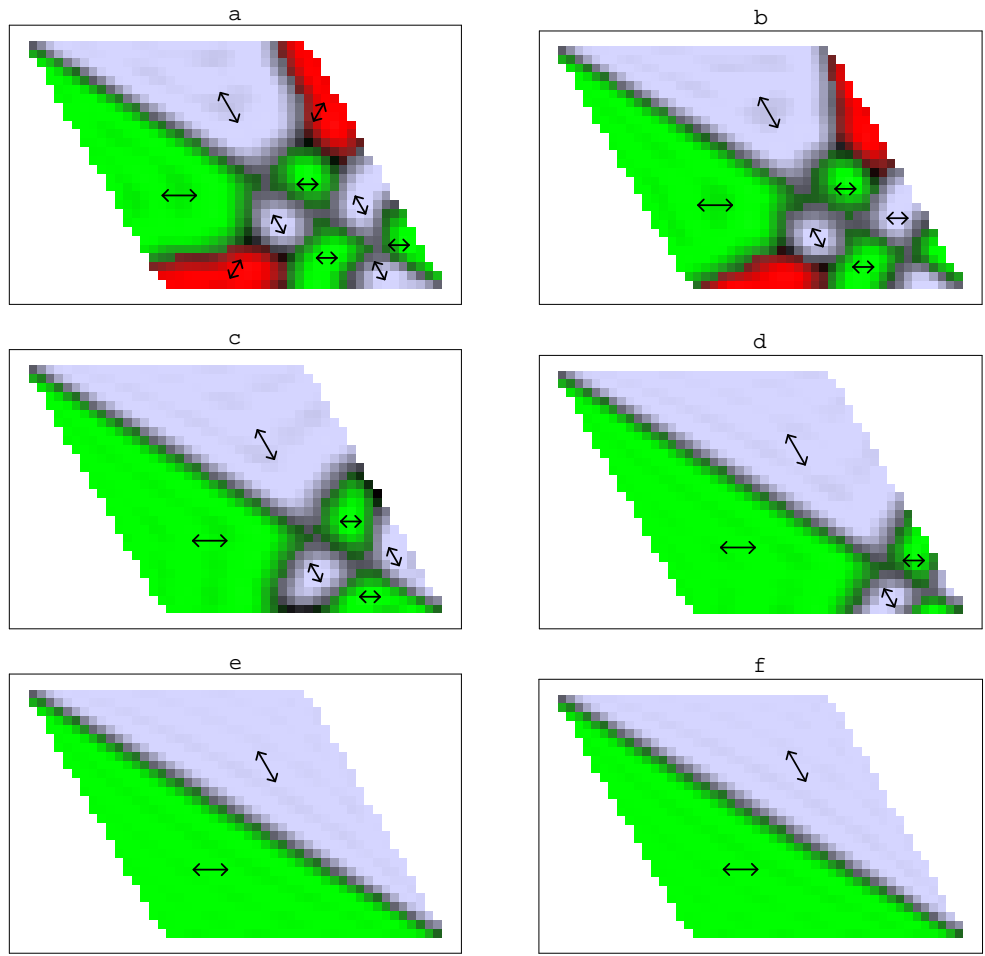,height=14cm}
\caption{Continuation of the series of snapshots from
Fig. 2, with the same spacing in time.  These panels show the
coarsening of the system at intermediate times in the dynamics,
during which the smaller domains shrink and are removed from the
system while the larger domains begin to domiante the 
system.  This figure also shows the reflection symmetry of 
the domain orientations, present in the initial condition, 
to good advantage.}
\label{fig:symd1}
\end{figure}

The phenomenon in which the initial patchwork of domains
all have approximately the same size scale is not universal
behaviour for the system, but rather is understood to be a
consequence of the interaction between
the inertial dynamics and the damping of the system.  The 
length--scale can in fact be quantitatively predicted from a simple
linear instability analysis [6].  We consider a travelling
shear--wave displacement field
of the form
\begin{equation}
u_x = \alpha e^{i(k\cdot y-\omega t)},~~~~~~~~~u_y = 0,
\end{equation}
for some amplitude $\alpha$.
Substituting this into the linear part of the Lagrangian,
Eq.~(\ref{eq:lag}), and into the dissipation function, Eq.~(\ref{eq:dis}),
we find that the resulting equation of motion gives
the frequency of the shear wave as a function of its
wavenumber.  The linear growth rate is the imaginary part of 
the frequency, and is given by 
\begin{equation}
{\rm Im}(\omega)={-\gamma_B \over 4\rho}k^2+
\left[ {|B| \over 8\rho}k^2-
   \left({g_2 \over 8\rho}-{\gamma_B^2 \over 16\rho^2}\right)
    k^4 \right]^{1 \over 2},
\label{eq:imomega}
\end{equation}
where we have chosen the most positive of the possible roots
of a quadratic relation in $\omega$.  The values chosen for
$\gamma_B$ and $g_2$ simplify the situation somewhat by
making the square root in Eq.~(\ref{eq:imomega}) purely imaginary.
The wavenumber with the maximum
growth rate is obtained by maximizing the above relation with respect 
to $k$, and we have
\begin{equation}
k^2_{\rm growth}=\left( |B| \over 
{2g_2+4\gamma_B \sqrt{g_2 \over 8\rho}}\right).
\end{equation}

Substituting the parameter values used for our dynamical 
study, we find 
that the length scale for the maximum growth rate is $l=0.7$, 
or approximately one--third of the length of one of the sides
of the diamond, and that the earliest fully--transformed
picture of the system supports the contention that a dynamically--selected
length scale of this magnitude dominates the morphology of the
early--stage dynamics.  This conclusion is further supported by
another run, not illustrated here [7], for which the parameter 
$B$ was set equal to zero, and a variety of lengths scales were
observed in the initially developed strain pattern.

The late stage of the dynamics, indicated in
Fig. 4, illustrates the importance of the 
reflection symmetry of the initial displacement field across
the long axis of the system.  This is a symmetry which is
preserved under the dynamics of the system, and which therefore
persists into the final state.  The only possible homogeneously 
strained minimum, and therefore the lowest possible energy state
for this system, is that in which the system is stretched along
the short axis.  The dynamics do not access this minimum, however,
and instead reaches a final state which consists of
two domains, consistent with the symmetry of the initial conditions,
and separated by a domain wall.  This patterned state
is expected to be arbitrarily long--lived under our
deterministic dynamics, although it is 
far from the lowest possible energy of the system.  

The model presented here incorporates several of the essential features
of diffusionless structural phase transformations which are known
to give rise to pattern formation in a variety of solid systems.
We have abstracted the essential features of such systems ---
symmetry, inertia, and dissipation --- which are sufficient to give
rise to the pattern--forming dynamics which are ubiquitous in 
a large variety of natural systems [8], while leaving out the
precise crystallographic details which vary between such systems and
which dominate them at much shorter length scales than those of
interest here.  Our abstract system is also not fully
three--dimensional, as are all real coherent nonlinear elastic
systems, but this is evidently not a requirement.

With only these essential features, and guided by the suggestive
results of the one--dimensional work, we have observed pattern--formation
dynamics, including one instance of an arbitrarily long--lived 
pattern.

We acknowledge helpful comments of Nick Schryvers and Bob Kohn. 
This work was supported by the NSERC of Canada.

\vfill
\eject
\end{document}